\documentclass[twocolumn,amsmath,amssymb]{revtex4}
\usepackage{graphicx}
\usepackage{dcolumn}
\usepackage{color}
\usepackage{bm}
\usepackage{epsfig}
\usepackage{enumerate}
\usepackage{array}
\usepackage{multirow}
\usepackage{hyperref}
\usepackage{siunitx}
\usepackage{scrextend}
\usepackage{subfig}
\usepackage{amsmath,amssymb,graphicx}
\usepackage{epsfig}
\usepackage{enumerate}
\usepackage{array}
\usepackage{multirow}
\begin{document}

\title{Stochastic thermodynamics and the Ericsson nano engine --\\ Efficiency from equilibrium results}

\author{J. Kaur$^1$, A. Ghosh$^1$\footnote{ag34@iitbbs.ac.in}, S. Dattagupta$^2$, S. Chaturvedi$^3$, M. Bandyopadhyay$^1$}

\affiliation{$^{1}$School of Basic Sciences, Indian Institute of Technology Bhubaneswar, Khurda, Odisha 752050, India\\
$^{2}$Sister Nivedita University, Kolkata, West Bengal 700156, India\\
$^{3}$Indian Institute of Science Education and Research, Bhopal, Madhya Pradesh 462066, India}

\vskip-2.8cm
\date{\today}
\vskip-0.9cm

\begin{abstract}
In this work, we study an Ericsson cycle whose working substance is a charged (quantum) oscillator in a magnetic field that is coupled to a heat bath. The resulting quantum Langevin equations with built-in noise terms encapsulate a thermodynamic structure and allow for the computation of the efficiency of the cycle. We numerically compute the efficiency of the cycle in the quasi-static regime using the steady-state thermodynamic functions of the system. Interestingly, it is found that by increasing the system-bath coupling strength, the efficiency of the cycle can be tuned to a maximum. We also explore the behavior of the efficiency as a function of the pair of magnetic-field values between which the cycle is operated.
\end{abstract}

\maketitle

\section{Introduction}\label{intro}
Since the demand for clean energy is increasing worldwide, there is a necessity for different kinds of technologies to provide reliable and sustainable energy to meet this demand. Small-scale external heat engines such as Ericsson engines effectively use different heat sources to perform work at the nanoscale. Although past research on the Ericsson engine has not been as substantial as that on the Stirling engine, yet there have been important developments in the former direction which are worth mentioning. Sisman and Saygin analyzed the efficiency of the Ericsson cycle for different working fluids \cite{sisman}, whereas, Bonnet and coworkers introduced an open-cycle Ericsson engine where the working fluid is constantly refurbished for micro-generation purposes \cite{bonnet}. Further, the relationships between the geometrical characteristics of the engine, its operating parameters, and its power and efficiencies were established by Tour\'e and Stouffs \cite{toure}. Finally, we should mention the work of Creyx and coworkers who analyzed a dynamic model and compared the effects of the air-intake pressure and different temperature conditions \cite{creyx}. In the present work, we shall be studying an Ericsson cycle where instead of a `fluid', the working substance is a trapped ion at finite temperature inside a harmonic well in the presence of an external magnetic field (see also, \cite{SC_SDG}); such a setup is expected to be realizable using modern-day experimental techniques.\\

Besides introducing a new kind of nanoscale Ericsson engine, our analysis is based on a contrasting viewpoint towards thermodynamics -- rather than considering the standard approach to equilibrium statistical mechanics which relies on the equilibrium density matrix or its normalizing partition function, we shall consider time-dependent equations of motion of observables that contain `dissipative' terms originating from the coupling to a surrounding heat bath. A particular example is that of the Langevin equations, which, with a built-in fluctuation-dissipation theorem, can yield ‘equilibrium’ results in agreement with statistical mechanics (see for instance, \cite{FordQT,hangpath,sdg3,malay1,kaur2}). The validity of the fluctuation-dissipation theorem rests on the assumption of ‘mixing’ which requires that all accessible points in the phase space are explored over an infinitely-long time \cite{ref2}. Appropriately, therefore, the latter ‘Brownian motion’ model is dubbed as the `Einstein approach to statistical mechanics' \cite{ref3,ref4} (see \cite{sdg3,malay1,SDG2} for some results on the present system of interest). The nice thing about this method is that in addition to equilibrium quantities, various non-equilibrium and `approach-to-equilibrium' properties can also be calculated.\\

In the recent years, the Einstein approach to statistical mechanics has been further extended to the domain of thermodynamics giving rise to what is called `stochastic thermodynamics' \cite{ref5,ref6} (see \cite{SDGStoc} for a simple and pedagogic introduction). The idea is to rewrite the Langevin equations by delineating the subsystem dynamics (including external-field-induced terms) from the bath-induced dissipative terms to put them in the context of the `differential’ heat, energy, and work `operators’ as in the first law of thermodynamics. Here, operators are written within quotes to emphasize that only when averaged over the noise terms inherent in the Langevin equations can they be ascribed thermodynamic interpretations. This approach is not only physically motivated but it also allows one to go beyond thermodynamics into the microscopic realm of fluctuating and time-dependent observables of the system. This method is also extremely useful in the topically-important applications to thermal ratchets, nano-Brownian motors, etc., especially in the context of classical biological processes \cite{ref6}. In this paper, we transit from the domain of classical to quantum phenomena and
assess stochastic thermodynamics of quantum Langevin equations as appropriate for dissipative quantum mechanics \cite{ref7,QLE,ref8} (see also, \cite{Zwanzig}). We present an exactly-solvable model of a charged quantum particle (such as an electron or an ion) in a two-dimensional parabolic well, subjected to a transverse magnetic field while in interaction with a heat bath that is taken to be composed of infinitely-many quantum oscillators \cite{ref9}. The problem of the dissipative cyclotron motion of an electron is of great interest in dissipative Landau diamagnetism and other condensed-matter issues such as the quantum Hall effect \cite{ref8}.\\

In the limit of what is called ‘Ohmic dissipation’ \cite{ref8}, we employ some known exact results to have a relook \textit{\`a la} Sekimoto \cite{ref6} at the first law of thermodynamics in which the magnetic field \(B\) plays the role of (negative) pressure \(P\), conjugate to the magnetization \(M\). With this identification, the main focus of the present paper is on computing the efficiency of the Ericsson cycle using steady-state results following an earlier development by two of the present authors \cite{SC_SDG} (see also, \cite{GSAW}). In particular, we emphasize upon the stochastic-thermodynamics approach to Brownian heat engines and demonstrate the transition between the strong and weak-coupling regimes. Quite interestingly, the approach based on the quantum Langevin equation, under a certain `weak-coupling' limit, can lead to the equilibrium results obtained from the master equation in the Born-Markov approximation \cite{GSA,kaur3,virial,agmb,AGSDG} (see also, \cite{weak1}). \\

Given the preceding remarks, we now present the outline of this paper. In the next section [Sec. (\ref{mmmsec})], we present the model of the dissipative magneto-oscillator \cite{ref9,kaur3}, wherein a charged quantum particle is placed in a magnetic field and a parabolic trap while in interaction with a bath composed of an infinitely-many quantum oscillators \cite{lho1,FV,CL,CL1,fordlho}. On integrating out the bath degrees of freedom, we quote the quantum Langevin equations (QLE) for the problem at hand viz., the dissipative cyclotron motion in a parabolic well. When the external magnetic field is taken to be varying, the QLE can be recast in the structure of the first law of thermodynamics. While the resultant stochastic thermodynamics discussed in Sec. (\ref{stocsec}) is capable of treating the full dynamics, we review in Sec. (\ref{steadysec}), known results that one obtains in the steady-state limit. We discuss quantities needed such as the heat, work, energy, and magnetization operators. Then, in Sec. (\ref{theorysetup}), we discuss the theoretical setup of the quantum Ericsson engine where the magneto-oscillator is made to perform work by changing the externally-applied magnetic field. The results of the ensuing numerical analysis are then presented in Sec. (\ref{numerics}). Finally, our principal conclusions are summarized in Sec. (\ref{conc}).\\

\section{Model}\label{mmmsec}
Let us consider a quantum particle of mass \(m\) and electric charge \(e\) confined in a harmonic potential with frequency \(\omega_0\). It is acted upon by a uniform magnetic field \(\mathbf{B}\), and it is bilinearly coupled to a heat bath which is composed of an infinite number of independent quantum oscillators, as in the independent-oscillator model \cite{fordlho}. The total Hamiltonian reads \cite{sdg3,malay1,SDG2,ref9}
\begin{eqnarray}
   H &=& \frac{(\mathbf{p} - \frac{e}{c} \mathbf{A})^2}{2m} + \frac{m \omega_0^2 \mathbf{r}^2}{2} \nonumber \\
   &&+ \sum_{j=1}^N\bigg[\frac{\mathbf{p}_j^2}{2m_j} + \frac{m_j \omega_j^2}{2} \bigg( \mathbf{q}_j - \frac{c_j}{m_j \omega_j^2}\mathbf{r} \bigg)^2 \bigg], \label{H}
\end{eqnarray}
where $\mathbf{p}$ and $\mathbf{r}$ are the momentum and position operators of the system, $\mathbf{p}_j$ and $\mathbf{q}_j$ are the corresponding operators for the \(j\)th  oscillator of the heat bath, and $\mathbf{A}$ is the vector potential, i.e., \(\mathbf{B} = \nabla \times \mathbf{A}\) is a constant vector. One has the usual commutation relations between coordinates and momenta. 
\begin{widetext}
Integrating out the heat-bath operators from the Heisenberg equations, one obtains a (vector) quantum Langevin equation (see \cite{ref8} for some details):
\begin{equation}\label{eqnm}
  m \ddot{\mathbf{r}}(t) + \int_{0}^{t} \mu(t - t') \dot{\mathbf{r}}(t') dt' + m \omega_0^2 \mathbf{r}(t)-\frac{e}{c}(\dot{\mathbf{r}}(t) \times \mathbf{B}) = \mathbf{f}(t),
\end{equation}
where \(\mu(t)\) is the dissipation kernel and \(\mathbf{f}(t)\) is the bath-induced noise which respectively read
\begin{equation}\label{mudef}
  \mu(t) = \sum_{j=1}^{N} \frac{c_j^2}{m_j \omega_j^2} \cos (\omega_j t) \Theta (t), \hspace{5mm}  \mathbf{f}(t)=\sum_{j=1}^{N} c_j \Bigg[\bigg(\mathbf{q}_j(0)-\frac{c_j}{m_j \omega_j^2}\mathbf{r}(0)\bigg)\cos(\omega_jt)+\frac{\mathbf{p}_j(0)}{m_j\omega_j}\sin(\omega_jt)\Bigg].
\end{equation} 
It should be noted that the noise depends upon the initial position of the system as well as the initial conditions of the heat-bath oscillators; these are taken to be distributed according to the canonical density matrix which goes as 
\begin{equation}\label{p0}
\rho_{\rm B + SB}(0) = Z^{-1} \exp\Bigg\{{-\beta \sum_{j=1}^N \bigg[ \frac{\mathbf{p}_j^2(0)}{2m_j} + \frac{m_j \omega_j^2}{2} \bigg(\mathbf{q}_j(0) - \frac{c_j}{m_j \omega_j^2} \mathbf{r}(0) \bigg)^2 \bigg]}\Bigg\}, \hspace{7mm} \beta = (k_B T)^{-1},
\end{equation} where \(Z\) is a suitable normalizing factor. Eq. (\ref{p0}) tells us that we have taken the system and the heat bath to be coupled at the initial instant, i.e., at \(t = 0\) \cite{bez,physicaA,SDG2}. This leads to non-trivial system-bath correlations even at the initial instant allowing us to go beyond the Born approximation which regards a `decoupled' initial preparation of the system and the bath \cite{AGSDG}. Now, with respect to this density matrix [Eq. (\ref{p0})], the noise is Gaussian, i.e., it has zero mean, all odd moments vanish, and the even moments can be written as ordered products of second moments. The spectral properties of the noise are characterized by the following symmetric correlation and commutator \cite{ref8,QLE}:
\begin{eqnarray}
\langle \lbrace f_{\rho}(t), f_{\sigma}(t') \rbrace \rangle &=& \frac{2 \delta_{\rho\sigma}}{\pi}\int_{0}^{\infty}d\omega \hbar \omega {\rm Re}[\tilde{\mu}(\omega)] \coth\bigg(\frac{\hbar\omega}{2k_BT}\bigg) \cos \lbrack \omega(t-t')\rbrack,  \label{symmetricnoisecorrelation} \\
\langle \lbrack f_{\rho}(t), f_{\sigma}(t') \rbrack\rangle &=& \frac{2\delta_{\rho\sigma}}{i\pi}\int_{0}^{\infty}d\omega \hbar \omega {\rm Re}[\tilde{\mu}(\omega)] \sin\lbrack \omega(t-t')\rbrack. \label{noisecommutator} 
\end{eqnarray} In the above-mentioned equations, \(\tilde{\mu}(\omega)\) represents the Fourier transform of the friction kernel \(\mu(t)\), and here $\sigma$ and $\rho$ are being used to indicate the Cartesian indices \(x\), \(y\), and \(z\). The angular brackets in Eqs. (\ref{symmetricnoisecorrelation}) and (\ref{noisecommutator}) imply thermal averaging over the heat bath, i.e., taking average over all noise realizations which basically implies averaging over the (initial) density matrix given in Eq. (\ref{p0}). \\

Let us recall that the bath spectral function \(J(\omega)\) characterizing the spectral distribution of the bath degrees of freedom is defined as \cite{ref7,ref8}
\begin{equation}\label{Jdef}
  J(\omega)=\frac{\pi}{2}\sum_{j=1}^{N} \frac{c_j^2}{m_j \omega_j} \delta(\omega-\omega_j) \hspace{2mm} \implies \hspace{2mm} \mu(t) = \frac{2}{\pi} \int_0^\infty \frac{J(\omega)}{\omega} \cos (\omega t) d\omega. 
\end{equation} 
\end{widetext}
In other words, the bath spectral function is related via cosine transform to the dissipation kernel \(\mu(t)\). Further, since the Fourier transform of \(\mu(t)\) appears in Eqs. (\ref{symmetricnoisecorrelation}) and (\ref{noisecommutator}), one can conclude that specifying the bath spectral function \(J(\omega)\) does indeed specify all the relevant details of the heat bath including the statistical properties of the noise. In this paper, we shall consider the case of Ohmic dissipation for which the low-frequency behavior of the bath spectral function reads as \(J(\omega) \sim \omega\) \cite{ref8}. Let us at the moment assume that the same linear dependence on frequency holds good for all frequencies. This gives \(\mu(t) \sim \delta(t)\), i.e., the `drag' force has no memory. This is the `strictly' Ohmic model of the heat bath and is somewhat controversial because it does lead to certain divergences \cite{grabert,grabert1} (see also, \cite{virial,agmb} and references therein). For instance, the variances of the momentum operators \(p_x\) and \(p_y\) diverge due to the divergent contributions coming from the zero-point energies of the heat-bath oscillators. In such situations, one usually employs a suitable regularization scheme to obtain a finite result \cite{grabert}. 

\subsection{Stochastic thermodynamics}\label{stocsec}
Let us now describe the framework of stochastic thermodynamics which shall motivate the subsequent study of the Ericsson cycle. We choose the magnetic field to be in the \(z\)-direction such that \(\mathbf{B} = B \hat{z}\), where \(B\) is a real constant. Then, the oscillator undergoes dissipative cyclotron motion on the \(xy\)-plane because of the Lorentz force that emerges due to the applied magnetic field \cite{ref9}. For the sake of simplicity, we will neglect the motion in the \(z\)-direction.
\begin{widetext}
The resulting two-dimensional Langevin equations take the following form: 
\begin{eqnarray}
{v}_x \equiv \dfrac{dx}{dt}&=& \frac{1}{m} p_x - \frac{1}{2}\omega_c y, \hspace{8mm} \frac{dp_x}{dt}= -\frac{1}{2}\omega_c p_y - m\left(\frac{\omega_c^2}{4} + \omega_0^2\right) {x} -
m\gamma v_x+ f_x, \nonumber \\
v_y\equiv \dfrac{dy}{dt}&=& \frac{1}{m} p_y + \frac{1}{2}\omega_c x, \hspace{8mm} \frac{dp_y}{dt}= ~~  \frac{1}{2}\omega_c p_x - m \left(\frac{\omega_c^2}{4} + \omega_0^2\right){y} - m\gamma {v}_y+ {f}_y, \label{qlemag}
\end{eqnarray} where we have implemented strict Ohmic dissipation which gives \(\mu(t) = 2 m \gamma \delta(t)\) for \(\gamma > 0\) and \(\omega_c = eB/mc\) is the cyclotron frequency. Here and later, for the sake of brevity, we shall suppress the argument \(t\) of the operators with the understanding that all the operators are evaluated at the 
same time, unless specified otherwise. The correlations of the noise operators appearing in Eqs. (\ref{symmetricnoisecorrelation}) and (\ref{noisecommutator}) in the case of Ohmic dissipation are given by 
\begin{eqnarray}
 \langle\{{f}_\rho(t), {f}_\sigma(t')\}\rangle &=& \delta_{\rho\sigma} \frac{2m\gamma}{\pi}\int_0^\infty d\omega \hbar
  \omega~\text{coth}\bigg(\frac{\hbar\omega}{2k_B T}\bigg) \cos[\omega(t-t')],  \\      
\langle[{f}_\rho(t), {f}_\sigma(t')]\rangle  &=&  \delta_{\rho\sigma} \frac{2m\gamma}{i\pi}\int_0^\infty 
d\omega \hbar \omega \sin [\omega(t-t')],
\label{16}
\end{eqnarray} where the indices \(\rho\) and \(\sigma\) correspond to the Cartesian indices \(x\) and \(y\) (we are neglecting the dynamics along \(z\)). 
\end{widetext}

The essential idea behind the setup of stochastic thermodynamics is to define appropriate heat, work, and energy operators that satisfy the first law of thermodynamics upon suitable averaging over the stochastic components of the dynamics \cite{ref5,SDGStoc,ref6}. Let us allow the magnetic field to be time-dependent, i.e., let us take \(B = B(t)\). Then, the time-derivative of the system Hamiltonian \(H_S\) (the first two terms of Eq. (\ref{H})) reads as
\begin{eqnarray}
\dfrac{dH_S}{dt} = \dfrac{m}{2} \dfrac{d}{dt} [({v}_x^2 + {v}_y^2)
+ \omega_0^2({x}^2 + {y}^2)]\nonumber\\
- \frac{e}{2c} ({y}{v}_x-{x}{v}_y) \dfrac{dB}{dt}, \label{17}
\end{eqnarray} where \(m v_{x,y} = p_{x,y} - \frac{e}{c} A_{x,y}\). Hence, following Sekimoto, the first law of thermodynamics can be written in the operator form as
\begin{equation}\label{FL}
d\mathcal{Q} = d\mathcal{E} +d\mathcal{W},  
\end{equation}
where the `differential' work operator reads
\begin{equation}
d\mathcal{W} \equiv -\mathcal{M} dB, \hspace{5mm}
\mathcal{M} \equiv \frac{e}{2c} ({y}{v}_x-{x}{v}_y), 
\label{19} 
\end{equation}
with $\mathcal{M}$ being the `magnetic moment' operator; further, 
\begin{equation}
d\mathcal{E} \equiv d( m[{v}_x^2 + {v}_y^2] /2 + m\omega_ 0^2[{x}^2 + {y}^2]/2), 
\label{20}
\end{equation}
is the `differential' energy operator, while
\begin{equation}
                               d\mathcal{Q} \equiv  (- m\gamma {v}_x + {f}_x) d{x} + (- m\gamma {v}_y + {f}_ y) d{y},  
                               \label{21}
                               \end{equation}
is the `differential' heat operator. Notice that because for an electron, \(e = - |e|\), the magnetic-moment operator defined in Eq. (\ref{19}) leads to the definition of magnetic moment as considered in \cite{ref9}. \\

Let us understand the picture that now emerges from Eq. (\ref{FL}). The first law as it appears in Eq. (\ref{FL}) now has a microscopic meaning in that all the relevant terms as defined in Eqs. (\ref{19})-(\ref{21}) allow for the calculation of 
not just their averages but the associated fluctuations and probability distributions as well, employing the statistics of the noise operators which are Gaussian but non-white processes.
While Eq. (\ref{qlemag}) can be solved exactly and the time-dependence of all relevant quantities can be formally determined, our aim here is to extract 
the asymptotic steady-state properties.

\subsection{Steady-state thermodynamic functions}\label{steadysec}
Let us now recall the steady-state results for the averaged energy and magnetic moment of the dissipative magneto-oscillator (see \cite{ref9,kaur2,kaur3,agmb} for more details). The thermally-averaged energy of the dissipative magneto-oscillator is obtained in the steady state as \cite{agmb}
\begin{equation}\label{E11}
E \equiv  \frac{m}{2} \lim_{t \rightarrow \infty} \langle ({v}_x(t)^2 + {v}_y(t)^2)
+ \omega_0^2({x}(t)^2 + {y}(t)^2) \rangle ,
\end{equation} while the steady-state magnetic moment reads \cite{ref9,kaur3}
\begin{equation}\label{M11}
M \equiv  \lim_{t \rightarrow \infty} -\frac{|e|}{2c} \langle {y}(t){v}_x(t)-{x}(t){v}_y(t)\rangle. 
\end{equation}
Recall that the angled brackets denote averaging over all-possible noise realizations, i.e., over the ensemble of noises. The corresponding expressions read as (see for example, \cite{ref9,kaur3})
\begin{eqnarray}
E &=&  \frac{1}{ \pi} \int_{-\infty}^\infty d\omega u(\omega)  (\omega^2 + \omega_0^2)  [\Phi(\omega) + \Phi(-\omega)] , 
\label{kineticenergysteadystate}\\
M &=& \frac{|e|}{ \pi m c}  \int_{-\infty}^\infty d\omega  \omega u(\omega) [\Phi(\omega) - \Phi(-\omega)] , 
\label{mz}
\end{eqnarray}
where \(u(\omega) = \frac{\hbar \omega}{4} \coth \big(\frac{ \hbar \omega}{2 k_B T} \big)\), and
\begin{eqnarray}
\Phi(\omega)= \frac{\gamma}{\big[\big(\omega^2-\omega_0^2-\omega\omega_c\big)^2+(\omega \gamma )^2\big]}.  \label{Phi2}
\end{eqnarray}
Evaluating the integrals above, one can determine the thermally-averaged energy and the magnetic moment exactly. The magnetic moment precisely reads \cite{ref9}
\begin{eqnarray}\label{mag22}
M &=& -2\frac{|e|}{ mc\beta}\sum_{n=1}^{\infty} \frac{\nu_n^2\omega_c}{[\nu_n^2+\omega_0^2+\gamma\nu_n]^2+(\nu_n\omega_c)^2},
\end{eqnarray}
where $\nu_n=\frac{2\pi n}{\hbar\beta}$ with $n= 1,2, \cdots $ being the (bosonic) Matsubara frequencies which arise from the poles of the coth function when expressed as
\begin{equation}
\zeta \coth \zeta = 1 + 2 \sum_{n=1}^\infty \frac{\zeta^2}{\zeta^2 + (n\pi)^2},
\end{equation} where \(\zeta\) is a general complex-valued argument. It may be remarked that in expressing the final result of Eq. (\ref{mag22}), we have put $\omega_c = |e|B/mc$ as opposed to the earlier definition given below Eq. (\ref{qlemag}). Turning now our attention to the thermally-averaged energy of the system, it follows that the resulting infinite series diverges \cite{agmb} because the terms for large \(n\) go as \(1/n\). Thus, let us impose a Lorentzian cutoff on the bath spectral function as
\begin{equation}
J(\omega) = \frac{m \gamma \omega \omega_{\rm cut}^2}{\omega^2 + \omega_{\rm cut}^2},
\end{equation} where \(\omega_{\rm cut}\) is a cutoff frequency which can be taken to be quite larger than the relevant frequency scale associated with the dissipative system, namely, \(\omega_0\). Then, the low-frequency behavior of the bath spectral function is still of the Ohmic type, i.e., \(J(\omega) \sim \omega\) for \(\omega << \omega_{\rm cut}\), while \(J(\omega)\) falls off for large values of frequency unlike the `strict' Ohmic case which is recovered for \(\omega_{\rm cut} \rightarrow \infty\). \\

\begin{widetext}
Now, for the present case with a finite cutoff, one can determine the thermally-averaged energy of the system in the form of an infinite series and the series is found to converge. Some straightforward manipulations give \cite{agmb}
\begin{equation}\label{E3dSeries}
E = 2k_B T \Bigg[1  +  \sum_{n=1}^{\infty}\frac{ \big( \nu_n^2 +\omega_0^2 + \frac{\nu_n\gamma \omega_{\rm cut}}{\nu_n +\omega_{\rm cut}}\big)\times \big( 2 \omega_0^2 + \frac{\nu_n\gamma\omega_{\rm cut}}{\nu_n +\omega_{\rm cut}}\big) + (\omega_c\nu_n)^2}{ \big( \nu_n^2 +\omega_0^2 + \frac{\nu_n\gamma \omega_{\rm cut}}{\nu_n +\omega_{\rm cut}}\big)^2+(\omega_c\nu_n)^2} \Bigg],
\end{equation} and one can clearly see that as \(\omega_{\rm cut} \rightarrow \infty\), the series diverges as \(1/n\) for large \(n\). Equipped now with the expressions for the magnetic moment and the thermally-averaged energy of the dissipative magneto-oscillator, one can attempt to analyze the Ericsson cycle. However, we shall first pause to discuss the relevant weak-coupling limit below. 
\end{widetext}

\subsection{Weak-coupling limit}\label{wkcoupling}
It is imperative to consider the weak-coupling limit (see also, \cite{agmb,AGSDG,weak1}) in which many of the calculations involving quantum master equations are performed due to the analytical tractability that exists in this limit. At the present outset, the weak-coupling limit is simply the limit \(\gamma << \omega_0, \omega_c\), i.e., we take the damping strength to be much smaller than all the relevant frequency scales of the problem. \\

For the magnetic moment of the system, the limit \(\gamma \rightarrow 0\) does coincide with \(\gamma = 0\) put into Eq. (\ref{mag22}). The resulting expression reads
\begin{equation}\label{dissipationzero}
M|_{\gamma \rightarrow 0}=-\frac{2B}{\beta}\Big(\frac{e}{mc}\Big)^2\sum_{n=1}^{\infty}\frac{\nu_n^2}{(\nu_n^2+\omega_0^2)^2+(\nu_n\omega_c)^2},
\end{equation} which can be regarded as the weak-dissipation expression for the magnetic moment of a charged oscillator when placed in a transverse and uniform magnetic field. Putting \(\omega_0 \rightarrow 0\) does consistently give rise to the result familiar from Landau's treatment of diamagnetism which reads
\begin{eqnarray}
M|_{\gamma, \omega_0 \rightarrow 0}&=&-\frac{2}{B\beta}\sum_{n=1}^{\infty}\frac{\omega_c^2}{(\nu_n^2+\omega_c^2)} \nonumber \\
&=& \frac{|e|\hbar}{2mc}\Big[\frac{2}{\beta\hbar\omega_c}-\coth\Big(\frac{\beta\hbar\omega_c}{2}\Big)\Big], \label{landauanswer}
\end{eqnarray} where the first term above corresponds to the surface contribution to the magnetic moment while the second term, i.e., that involving the coth function gives the bulk contribution \cite{ref9,kaur3}. We shall, however, set \(\omega_0\) to some finite (non-zero) value as it sets the relevant frequency scale of the system; in the subsequent numerical analysis, we shall express the numerical values of the other parameters in units of \(\omega_0\). 

\begin{widetext}
Considering now the thermally-averaged energy of the dissipative magneto-oscillator as given in Eq. (\ref{E3dSeries}), putting \(\gamma \rightarrow 0\) gives
\begin{eqnarray}
E|_{\gamma \rightarrow 0} = 2k_B T \Bigg[1  +  \sum_{n=1}^{\infty}\frac{ 2 \omega_0^2 \big( \nu_n^2 +\omega_0^2\big) + \big( \omega_c\nu_n \big)^2}{ \big( \nu_n^2 +\omega_0^2\big)^2+(\omega_c\nu_n)^2} \Bigg]. 
\label{Egammazero}
\end{eqnarray} 

It is curious to inquire about the possibility of taking \(\gamma \rightarrow 0\) in Eqs. (\ref{kineticenergysteadystate}) and (\ref{mz}) prior to performing the contour integrals. For this we recall the following well-known result:
\begin{equation}\label{deltafunction}
\pi \delta(x - x_0) =  \lim_{a \rightarrow 0^+} \frac{a}{(x-x_0)^2 + a^2},
\end{equation} which means, putting \(\gamma \rightarrow 0\) in Eq. (\ref{kineticenergysteadystate}) gives
\begin{eqnarray}
E|_{\gamma \rightarrow 0} = \frac{\hbar \omega_1}{2} \coth \bigg(\frac{\hbar \omega_1}{2 k_B T}\bigg) + \frac{\hbar \omega_2}{2} \coth \bigg(\frac{\hbar \omega_2}{2 k_B T}\bigg), \hspace{7mm} \omega_{1,2} = \frac{1}{2} (\sqrt{4\omega_0^2 + \omega_c^2} \pm \omega_c ).
\label{Egammazero11}
\end{eqnarray}
 Here, \(\omega_{1,2}\) are the normal-mode frequencies of a charged  two-dimensional oscillator in the presence of a transverse magnetic field \cite{kaur3,malay2DEG}; putting \(\omega_c = 0\) gives \(\omega_{1,2} = \omega_0\). It may be verified that Eq. (\ref{Egammazero11}) is identical to Eq. (\ref{Egammazero}). Further, putting \(\omega_0 = 0\) gives \(\omega_1 = \omega_c\) and \(\omega_2 = 0\), leading to
 \begin{equation}
 E|_{\gamma,\omega_0 \rightarrow 0} = \frac{\hbar \omega_c}{2} \coth \bigg(\frac{\hbar \omega_c}{2 k_B T}\bigg),
 \end{equation} which is precisely the thermally-averaged energy of a one-dimensional quantum oscillator with frequency \(\omega_c\). It is noteworthy that as far as the energy is concerned, taking \(\gamma \rightarrow 0\) either in Eq. (\ref{kineticenergysteadystate}) or in Eq. (\ref{E3dSeries}) gives the same result. \\
 
However, taking the limit \(\gamma \rightarrow 0\) in Eq. (\ref{mz}) and performing the integral (utilizing Eq. (\ref{deltafunction})) gives
 \begin{equation}
 M|_{\gamma \rightarrow 0} = \frac{-|e|\hbar}{2mc} \bigg(\frac{1}{\omega_1 + \omega_2}\bigg) \bigg[ \omega_1 \coth \bigg(\frac{\hbar \omega_1}{2 k_B T} \bigg) - \omega_2 \coth \bigg(\frac{\hbar \omega_2}{2 k_B T}\bigg) \bigg]. 
 \end{equation}
Putting \(\omega_0 = 0\) gives \(\omega_1 = \omega_c\) and \(\omega_2 = 0\), and therefore the above-mentioned expression only recovers the bulk part of the magnetic moment, i.e., the second term inside the parenthesis of Eq. (\ref{landauanswer}). This highlights the crucial function played by the ordering of limits in the context of diamagnetism, where the boundary (set due to \(\omega_0\)) plays an important role in determining the magnetic moment \cite{ref9,P}. Setting \(\gamma \rightarrow 0\) in Eqs. (\ref{mz}) and (\ref{mag22}) does not give the same final result; one has to perform the integral over \(\omega\) first while also having \(\omega_0 \neq 0\) which takes into account the effect of the boundary. Reversing this order washes away the boundary contribution and one lands up with just the bulk part of the magnetic moment. 
\end{widetext}

\section{Quantum Ericsson cycle: Theoretical Setup}\label{theorysetup}
In this section, we will describe the Ericsson cycle where the dissipative magneto-oscillator acts as the working substance (see also, the preceding work \cite{SC_SDG}). We shall then investigate the efficiency of the cycle in Sec. (\ref{numerics}) using steady-state (equilibrium) results. Let us remind the reader that the Ericsson cycle for a hydrostatic system is a 4-stroke cycle consisting of two isothermal `legs' joined by two isobaric `legs'. In the present situation, referring to Eq. (\ref{17}), we find that it is the magnetic field that is analogous to the pressure of a hydrostatic system (up to a negative sign), meaning that an `isobaric' process corresponds to that with a fixed magnetic field or equivalently fixed \(\omega_c\). Motivated by this, we consider the following cycle: 
\noindent
\begin{align} \nonumber
&\begin{array}{cccccc}
B_1,T_h~&  &\text{Isothermal}& &&B_2,T_h\\
& 1 &\longrightarrow& 2&&\\
&   &&\\
 \text{Iso-\(B\)}&\uparrow&&\downarrow&&\text{Iso-\(B\)}\\
&   &&\\
& 4&\longleftarrow&3 &&\\
B_1,T_c~& &\text{Isothermal}& &&B_2,T_c\\
&&&&&\\ \end{array} \\
&~~~~~~~~~B_2>B_1,~~~~T_h>T_c~~~~~~~~~~
\nonumber
\end{align}

The magneto-oscillator starts with thermal equilibrium at temperature \(T_h\) and with a magnetic field \(B_1\) at point `1'. It is then taken to point `2' isothermally by changing the magnetic field to a new value \(B_2\). Such a change is taken to happen as
\begin{equation}\label{protocol}
B(t)  = B_1 + (B_2 - B_1) \frac{t}{\tau},
\end{equation} such that \(B(0) = B_1\) and \(B(\tau) = B_2\). Here, \(\tau\) is the timescale that characterizes this change and for simplicity, we shall take this to be much longer than the relaxation timescale of the system \footnote{It is important to remark here that the precise choice of a large-enough \(\tau\) does not significantly alter the efficiency of the cycle in the numerical results; for example, the choices \(\tau = 10^3\) and \(\tau = 10^4\) give almost identical results.}. Therefore, in effect, we can neglect the transient effects to a first approximation, and use the results presented in Sec. (\ref{steadysec}) to compute the efficiency of the cycle. The work done is computed from the (differential) work operator appearing in Eq. (\ref{19}) by averaging over the noisy effects inherent in the quantum Langevin equations and then integrating the infinitesimal work along the path. Thus, we have for the work done, the following expression:
\begin{equation}
W_{1 \rightarrow 2} = - \int_1^2 M dB,
\end{equation} where \(M\) is the steady-state result as it appears in Eq. (\ref{mag22}) with temperature \(T_h\). Further, the integral above leading to the work done has to be evaluated along the protocol given in Eq. (\ref{protocol}). Notice here that work is performed by changing the value of the magnetic field which neither affects the heat bath, nor does it affect the system-bath interaction; this ensures the consistency of the definition of work as given above \cite{hanggiRMP}.\\

Next, consider the process from `2' to `3' which takes place as an iso-\(B\) process. The total work done in this part of the cycle vanishes naturally, i.e., \(W_{2 \rightarrow 3} = 0\). The remaining two legs of the cycle, namely, `3' to `4' and `4' to `1' are performed in a similar way as the ones that we have already discussed. The total work done in a cycle is therefore \(W = W_{1 \rightarrow 2} + W_{3 \rightarrow 4}\). The efficiency of the cycle is given by the usual definition that goes as
\begin{equation}
\eta = \frac{W}{Q},
\end{equation} where \(Q\) is the net heat (input) given to the oscillator and can be decomposed as \(Q = Q_{1 \rightarrow 2} + Q_{4 \rightarrow 1}\). That the magneto-oscillator does no work along the leg \({4 \rightarrow 1}\) leads to the fact that \(Q_{4 \rightarrow 1} = \Delta E_{4 \rightarrow 1}\), where \(\Delta E_{4 \rightarrow 1}\) is the corresponding change in the thermally-averaged energy of the system. Furthermore, from the first law, we must have \(Q_{1 \rightarrow 2} = W_{1 \rightarrow 2} + \Delta E_{1 \rightarrow 2}\), where symbols have their usual meanings. Putting all this together, we can find the efficiency of the cycle as
\begin{equation}
\eta = \frac{W_{1 \rightarrow 2} + W_{3 \rightarrow 4}}{W_{1 \rightarrow 2} + \Delta E_{1 \rightarrow 2} +  \Delta E_{4 \rightarrow 1}}.
\end{equation}
In what follows, we estimate the efficiency numerically.

\section{Numerical Results}\label{numerics}
In the numerical results presented below, we take \(\tau = 1000\), i.e., the isothermal `expansion' is performed over a thousand discrete steps as Eq. (\ref{protocol}) suggests. Furthermore, we shall take \(T_h = 5 T_c\) which means the corresponding Carnot efficiency reads \(\eta_{\rm C} \equiv 1 - \frac{T_c}{T_h} = 0.8\), setting an upper bound on the maximum efficiency which may be achieved. We also maintain \(\omega_{c,2} = 10 \omega_{c,1}\) throughout all the computations. We take \(\hbar = k_B =1\), and measure all energy scales with respect to \(\omega_0\). \\

It needs to be pointed out that one is required to pick a minimum temperature for the `cold' bath as for extremely-low temperatures, the validity of the Clausius inequality, i.e., \(dQ \leq TdS\) may become dubious \cite{claus} and the computed efficiency may exceed that of the Carnot cycle. Moreover, for numerical computations involving expressions containing Matsubara sums (for example, Eq. (\ref{mag22})), there are subtleties involving taking the temperature to be arbitrarily close to zero as in that case the Matsubara frequencies come arbitrarily close to each other and the sum converts to an integral which often requires further regularization. For our purposes, we take \(T_{c,{\rm mininum}} = 0.002 \omega_0\).

\subsection{Weak system-bath coupling}
\begin{figure}
\begin{center}
\includegraphics[scale=0.90]{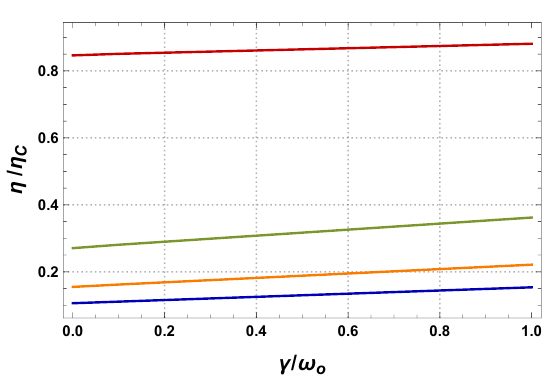}
\caption{Variation of efficiency (ratio with Carnot efficiency) \(\eta/\eta_C\) as a function of \(\gamma/\omega_0\), for \(T_c/\omega_0 =\) 0.002 (red), 0.008 (green), 0.014 (orange), and 0.02 (blue). We have taken \(\omega_{c,1}/\omega_0 = 0.1\), and for computing the energy differences, we have used \(\omega_{\rm cut}/\omega_0 = 10\).}
\label{fig1}
\end{center}
\end{figure}

\begin{figure}
\begin{center}
\includegraphics[scale=0.90]{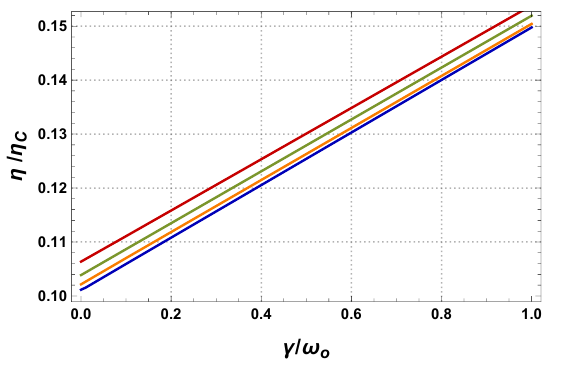}
\caption{Variation of efficiency (ratio with Carnot efficiency) \(\eta/\eta_C\) as a function of \(\gamma/\omega_0\) for \(\omega_{c,1}/\omega_0 = 0.1\) (red),  0.07 (green), 0.04 (orange), and 0.01 (blue). We have taken \(T_c/\omega_0 = 0.02\), and for computing the energy differences, we have used \(\omega_{\rm cut}/\omega_0 = 10\).}
\label{fig2}
\end{center}
\end{figure}

Let us first present the results for the efficiency emphasizing upon the weak-coupling limit which is achieved as \(\gamma << \omega_0, \omega_c\). We have plotted the efficiency of the cycle as a function of the rescaled damping strength \(\gamma/\omega_0 \leq 1\) in Fig. (\ref{fig1}) for different values of \(T_c\) with the smallest value for red (\(T_c/\omega_0 = 0.002\)) and the largest value corresponds to blue (\(T_c/\omega_0 = 0.02\)). Surely then, the efficiency of the cycle goes up at lower temperatures while it goes down as the cycle is operated between heat baths at higher temperatures; this makes perfect sense when one realizes that the magnetic Ericsson cycle has no classical counterpart for the Bohr-van Leeuwen theorem indicates that the magnetization must vanish at high temperatures and so does the work done by the cycle leading to decreased efficiencies as the temperature is increased, finally reaching the classical limit where the efficiency vanishes. Next, we have plotted the efficiency of the cycle as a function of \(\gamma/\omega_0 \leq 1\) for different values of the magnetic fields (keeping \(\omega_{c,2} = 10\omega_{c,1}\)) in Fig. (\ref{fig2}). The magnetic fields are the largest for red (\(\omega_{c,1}/\omega_0 = 0.1\)) and the smallest for blue (\(\omega_{c,1}/\omega_0 = 0.01\)) indicating towards the fact that the cycle performs more efficiently under larger magnetic fields. \\

Coming now to the limit \(\gamma \rightarrow 0\), i.e., \(\gamma << \omega_0, \omega_c\); it so turns out that the equilibrium attributes such as the magnetic moment or the thermally-averaged energy of the magneto-oscillator are independent of \(\gamma\) [Eqs. (\ref{dissipationzero}) and (\ref{Egammazero})]. This is a well-known feature of equilibrium thermodynamics: the attributes become independent of the system-bath coupling strength when the latter is considered to be sufficiently `small'. For our present purposes, we use the equilibrium results to find the efficiency which therefore has no dependence on the system-bath coupling. We have verified  that upon taking \(\gamma << \omega_0,\omega_c\), the efficiency computed in general reduces to that obtained independently in the weak-coupling limit using Eqs. (\ref{dissipationzero}) and (\ref{Egammazero}) (the \(y\)-intercept in Figs. (\ref{fig1}) and (\ref{fig2})). 

\subsection{Larger values of system-bath coupling}
We verified that the results obtained for arbitrary coupling strengths do reduce to the weak-coupling results as one takes \(\gamma \rightarrow 0\) in the former calculations. In fact, the damping strength \(\gamma\) seems to have significant control over the efficiency of the cycle. Now, although from Figs. (\ref{fig1}) and (\ref{fig2}), one finds that the efficiency of the cycle increases with the system-bath coupling strength, it cannot increase indefinitely and monotonically for the efficiency cannot exceed the Carnot efficiency. In fact, what one expects is that the efficiency should decrease for large-enough values of the damping strength as bath-induced decoherence dominates over the coherent motion of the charged particle; the latter leads to diamagnetism and the former destroys it \cite{cursci}. This leads to a bath-induced quantum-classical crossover in the sense that the magnetic moment decreases and goes to zero for large values of the damping strength, thereby restoring the classical `Bohr-van Leeuwen'-like result even within the quantum regime \((\hbar \neq 0)\) \cite{kaur3}. \\

\begin{figure}
\begin{center}
\includegraphics[scale=0.90]{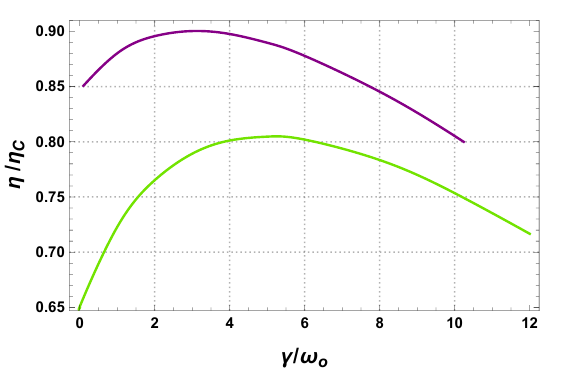}
\caption{Behavior of efficiency (ratio with Carnot efficiency) \(\eta/\eta_C\) as a function of \(\gamma/\omega_0\) for \(T_c/\omega_0 = 0.002\) (purple) and \(0.003\) (green). We have used \(\omega_{c,1}/\omega_0 = 0.1\), and for computing the energy differences, we have used \(\omega_{\rm cut}/\omega_0 = 1000 \).}
\label{fig5}
\end{center}
\end{figure}

\begin{figure}
\begin{center}
\includegraphics[scale=0.90]{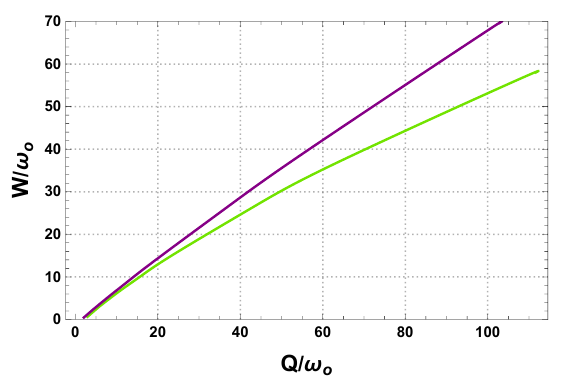}
\caption{Variation of rescaled work done \(W/\omega_0\) as a function of the rescaled heat input \(Q/\omega_0\) for \(T_c/\omega_0 = 0.002\) (purple) and \(0.003\) (green). We have used \(\omega_{c,1}/\omega_0 = 0.1 \), and for computing the energy differences, we have used \(\omega_{\rm cut}/\omega_0 = 1000\).}
\label{fig6}
\end{center}
\end{figure}

In Fig. (\ref{fig5}), we have plotted the efficiency as a function of the rescaled damping strength \(\gamma/\omega_0\), extending well beyond the regime \(\gamma \leq \omega_0\) which was considered earlier. It is found that the efficiency does increase from its weak-coupling value up to a certain point, after which any further increase in the damping strength does lead to a fall in the efficiency. In this sense, one may view the damping strength to be a control parameter that may be tuned to make the Ericsson cycle operate with the largest-possible efficiency between a given pair of temperatures \((T_c,T_h)\) and magnetic-field parameters \((\omega_{c,1},\omega_{c,2})\). In Fig. (\ref{fig6}), we have plotted (in units of \(\omega_0\)), the work done by the engine and the heat input over a single cycle corresponding to the case presented in Fig. (\ref{fig5}). \\

\begin{figure}
\begin{center}
\includegraphics[scale=0.90]{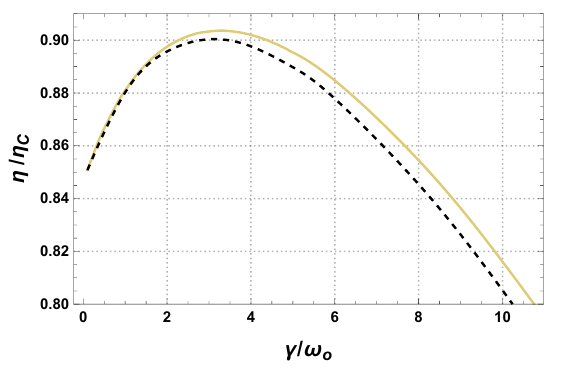}
\caption{Behavior of efficiency (ratio with Carnot efficiency) \(\eta/\eta_C\) as a function of \(\gamma/\omega_0\) for \(T_c/\omega_0 = 0.002\) and \(\omega_{c,1}/\omega_0 = 0.1 \), where, for computing the energy differences, we have used \(\omega_{\rm cut}/\omega_0 = 10\) (yellow) and \(\omega_{\rm cut}/\omega_0 = 1000 \) (black-dashed).}
\label{fig7}
\end{center}
\end{figure}

It should be remarked here that in Figs. (\ref{fig5}) and (\ref{fig6}), we have used \(\omega_{\rm cut}/\omega_0 = 1000 \) for computing energy differences to ensure that the cutoff-frequency scale is much larger than all the other energy scales associated with the dissipative system. In Fig. (\ref{fig7}), we have plotted the behavior of the efficiency as a function of the rescaled damping strength \(\gamma/\omega_0\) for the choices \(\omega_{\rm cut}/\omega_0 = 10 \) and \(\omega_{\rm cut}/\omega_0 = 1000\) for a particular choice of the other parameters; both the cases exhibit the same qualitative behavior and give (almost) identical results for smaller values of \(\gamma\). Therefore, it is the underdamped regime (typically denoted by \(\gamma < 2\omega_0\)) in which our results are the most reliable, first, due to relative insensitivity to \(\omega_{\rm cut}\) (notice that the magnetic moment is computed by assuming that \(\omega_{\rm cut} >> \omega_0,\omega_c,\gamma\)), and second, due to subtleties associated with defining heat in the regime with strong system-bath coupling \cite{hanggiRMP} which we do not address here. 

\subsection{Effect of strong magnetic fields}
\begin{figure}
\subfloat[\(T_c/\omega_0 = 0.002\).\label{fig:test1}]
  {\includegraphics[width=0.98\linewidth]{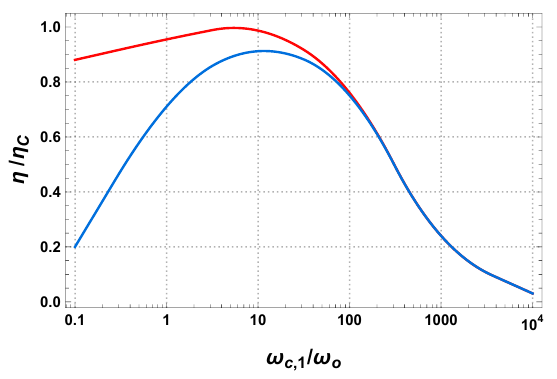}}\hfill
\subfloat[\(T_c/\omega_0 = 0.02\).\label{fig:test2}]
  {\includegraphics[width=0.98\linewidth]{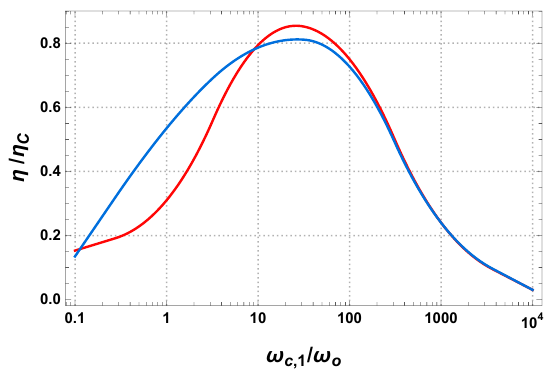}}
\caption{Variation of efficiency (ratio with Carnot efficiency) \(\eta/\eta_C\) as a function of \(\omega_{c,1}/\omega_0\) for \(\gamma/\omega_0 = 1\) (red) and \(\gamma/\omega_0 = 50\) (blue), for two different choices for \(T_c\) (\(T_h = 5T_c\)). For computing the energy differences, we have used \(\omega_{\rm cut}/\omega_0 = 10^6\).}
\end{figure}

Since the magnetic field plays a central role in controlling the steady-state behavior of the dissipative magneto-oscillator \cite{ref9,SC_SDG,kaur3,agmb,SDG2,kaur2,cursci,sdg3}, it seems worthwhile to investigate the effect of the strength of the magnetic-field parameters \(\omega_{c,1}\) and \(\omega_{c,2}\) (such that \(\omega_{c,2} = 10 \omega_{c,1}\)) on the efficiency of the Ericsson cycle. In Figs. (\ref{fig:test1}) and (\ref{fig:test2}), we have plotted the efficiency as a function of \(\omega_{c,1}/\omega_0\) for two different values of the system-bath coupling strength (and two different temperature choices) from which one observes a similar qualitative behavior in all the cases. It is found that the efficiency first increases with the magnetic field, reaches a maximum, and then decreases with further increase in the strength of the applied magnetic fields between which the cycle operates. For a given choice of parameters, there seems to exist an optimum value for the applied magnetic fields for which the cycle admits maximum efficiency.

\section{Closing remarks}\label{conc}
In this paper, we have investigated the efficiency of the Ericsson cycle in which a dissipative magneto-oscillator acts as the working substance and work is performed by changing the magnetic field applied to the system. Equilibrium results were employed to numerically estimate the efficiency of the cycle and its dependence on the various parameters. It was found that the efficiency is a non-monotonic function of the system-bath coupling strength (also called damping strength). It was also observed that the efficiency of the cycle decreases as one operates it between heat baths at higher temperatures. \\

As we conclude this paper, it should be pointed out that in any realistic operation of such a heat engine, one should expect to encounter finite-time effects (see for example, \cite{GSAW}) which could, in principle, be tackled using the framework of stochastic thermodynamics which is based on quantum Langevin equations. In particular, a realistic operation of the cycle shall involve decoupling from and coupling to heat baths which would lead to important transient effects. Thus, the present study which involves the idealized case may be regarded only as an initial step towards understanding the magnetic Ericsson cycle, which, we hope will be explored further in future publications.

\section*{Acknowledgements}
We thank Shamik Gupta for collaboration during early stages of this project. J.K. is grateful to Sibasish Ghosh for some useful remarks and thanks IIT Bhubaneswar for financial support in the form of an Institute Fellowship. A.G. acknowledges useful discussions with Peter Talkner and Gert-Ludwig Ingold. The work of A.G. is supported by the Ministry of Education (MoE), Government of India in the form of a Prime Minister's Research Fellowship (ID: 1200454). S.D. thanks the Indian National Science Academy for support through their Honorary Scientist Scheme. M.B. is supported by the Department of Science and Technology (DST), Government of India under SERB-MATRICS scheme Grant No. MTR/2021/000566.

\end{document}